\documentclass[
 reprint,
 superscriptaddress,
 amsmath,amssymb,
 aps,
 prx,
longbibliography]{revtex4-2}

\usepackage{graphicx}
\usepackage{dcolumn}
\usepackage{bm}
\usepackage{color}
\usepackage{hyperref}
\hypersetup{
	colorlinks = true,
	linkcolor = black,
	urlcolor = blue,
	citecolor = blue
}

\usepackage{placeins} 

\definecolor{orange}{rgb}{.9,0.45,0}

\definecolor{green}{rgb}{0,0.6,0} 

\newcommand\redout{\bgroup\markoverwith{\textcolor{red}{\rule[.5ex]{2pt}{1.2pt}}}\ULon}
\newcommand\blueout{\bgroup\markoverwith{\textcolor{blue}{\rule[.5ex]{2pt}{1.2pt}}}\ULon}

\begin{document}

\preprint{APS/123-QED}

\title{Experimental Quantum Tomography of Multimode Gaussian states}

\author{Chan Roh}
\affiliation{Department of Physics, Korea Advanced Institute of Science and Technology, Daejeon 34141, Korea}

\author{Geunhee Gwak}
\affiliation{Department of Physics, Korea Advanced Institute of Science and Technology, Daejeon 34141, Korea}

\author{Young-Do Yoon}
\affiliation{Department of Physics, Korea Advanced Institute of Science and Technology, Daejeon 34141, Korea}

\author{Young-Sik Ra}
 \email{youngsikra@gmail.com}
\affiliation{Department of Physics, Korea Advanced Institute of Science and Technology, Daejeon 34141, Korea}

\date{\today}

\begin{abstract}

Multimode Gaussian states are a versatile resource for quantum information technologies and have been realized across a wide range of physical platforms. Recent progress in the large-scale generation of such states provides a key ingredient for scalable quantum technologies. Despite the importance of accurately characterizing these states, conventional tomography methods are often impractical because they require large sample sizes and can yield unphysical states. Here we present a reliable and efficient tomography method for multimode Gaussian states based on maximum-likelihood estimation. By directly operating on covariance matrices, the method avoids the exponential overhead associated with density-matrix reconstruction. We consider two commonly used detection schemes---single and joint homodyne detection---and systematically analyze the reconstruction performance. Our method outperforms conventional approaches by ensuring physical covariance matrices and achieving better agreement with the true states. To demonstrate the experimental applicability of the method, we experimentally generate various multipartite entangled states---six-mode graph states with different connectivity, a six-mode GHZ state, and a fully connected ten-mode graph state---and reconstruct their covariance matrices. Using the reconstructed covariance matrices, we quantify fidelities, detect entanglement, and reveal the multimode structure of squeezing and noise. Our technique offers a practical diagnostic tool for developing scalable quantum technologies.

\end{abstract}

\maketitle


\section{Introduction}

Multimode Gaussian states play a central role in continuous-variable quantum information processing. These states naturally arise in various physical platforms, including photons~\cite{Larsen19,Asavanant19,Roh25,Barakat2025}, trapped ions~\cite{Chen2023,Li25}, mechanical oscillators~\cite{Kotler2021,Jeong25}, and superconducting microwave circuits~\cite{Mallet2011, Lingua2025}. Recent experimental advances in the generation of large-scale entangled Gaussian states provide a promising route toward scalable quantum technologies~\cite{Larsen19,Asavanant19,Roh25,Barakat2025,Lingua2025}. Examples include measurement-based quantum computing~\cite{Larsen19,Asavanant19,Roh25,Lingua2025}, Gaussian boson sampling~\cite{Zhong20,Madsen22}, quantum networks~\cite{Kovalenko21,Shi2023}, and multiparameter quantum metrology~\cite{Guo20,Barbieri:2022hq}. These technologies require precise knowledge of the underlying quantum states, making quantum state tomography essential~\cite{James01,Lvovsky09}.

Quantum state tomography for continuous-variable systems typically reconstructs the density matrix in the photon-number basis~\cite{Lvovsky04,Chapman22}. However, the number of independent parameters in the density matrix grows exponentially with the number of modes, rendering multimode tomography challenging~\cite{Lvovsky04,Ra:2020gg}. Many efforts have been made to develop efficient tomography methods. Compressed sensing assumes a low rank density matrix, yet it still suffers from exponential scaling~\cite{Gross10}. Restricting the total photon number can reduce the effective dimension of the density matrix~\cite{He24}, but this approach is not suitable for Gaussian states. Machine-learning approaches have also been explored, although they typically rely on training over a restricted set of states~\cite{Tiunov20,Hsieh:2022jc}.

Alternatively, multimode Gaussian states can be fully characterized by the first moments and the covariance matrix of the quadrature operators~\cite{Weedbrook12}. In particular, the covariance matrix captures all quantum properties of multimode Gaussian states~\cite{Weedbrook12,Serafini2017}. Importantly, the number of independent parameters scales quadratically with the number of modes~\cite{Serafini2017}, avoiding the exponential scaling of density-matrix reconstruction. Moreover, the sample complexity for learning Gaussian states scales polynomially with the number of modes~\cite{Mele25}. A conventional method for reconstructing the covariance matrix is to directly compute covariances from quadrature outcomes obtained via homodyne detection. Using this method, the full covariance matrix of a two-mode Gaussian state has been reconstructed~\cite{D'Auria09}, while partial covariance matrices (excluding $\hat{x}$-$\hat{p}$ correlation parts) have been reconstructed for Gaussian states in eight modes~\cite{Roslund14} and sixteen modes~\cite{Cai17}.

However, the conventional reconstruction method often yields unphysical covariance matrices due to statistical noise. Such unphysical matrices violate the uncertainty relation~\cite{Weedbrook12}, making them inapplicable for further analyses such as fidelity calculation~\cite{Banchi15}, entanglement detection~\cite{Simon00}, and quantum-steering quantification~\cite{Kogias15}. Several approaches have been used to mitigate this issue, including selecting only physical covariance matrices~\cite{Mele25}, adding white noise~\cite{Gerke15}, and projecting onto the closest physical covariance matrix~\cite{Shchukin24}. More recently, systematic approaches based on maximum-likelihood estimation (MLE) have begun to be explored~\cite{Roh25,Gwak25}, although their performance for multimode Gaussian states has not yet been systematically investigated.

Here we present MLE-based reconstruction of physical covariance matrices for multimode Gaussian states. We consider two informationally complete measurements: single homodyne detection and joint homodyne detection. We analyze the reconstruction performance of the MLE method compared with the conventional method. Specifically, we examine the physicality of the reconstructed matrices, evaluate their fidelity to the true covariance matrices, and assess their entanglement properties. Furthermore, we conduct experiments applying the method to various multimode Gaussian states, including six-mode graph states with different connectivity, a six-mode GHZ state, and a fully connected ten-mode graph state. We analyze the reconstructed covariance matrices by quantifying the fidelity, detecting entanglement, and revealing the multimode structure of squeezing and noise.

\section{Gaussian state and measurement}

\subsection{Description of multimode Gaussian states}

We begin by establishing a general description of multimode Gaussian states with $M$ modes. The amplitude and phase quadrature operators of $m$-th mode are defined as $\hat{x}_m = \hat{a}_m+\hat{a}_m^\dagger$ and $\hat{p}_m = i(\hat{a}_m^\dagger - \hat{a}_m)$, respectively, where $\hat{a}_m$ ($\hat{a}_m^\dagger$) denotes the annihilation (creation) operator of the $m$-th mode. For compactness, we introduce the quadrature operator vector, $\hat{\boldsymbol{q}} = [\hat{x}_1,...,\hat{x}_M,\hat{p}_1,...,\hat{p}_M]^T$. The corresponding commutation relations are given by $[\hat{q}_m, \hat{q}_n] = 2 i \Omega_{mn}$, where the symplectic form is
\begin{equation}
	\boldsymbol{\Omega} = \begin{bmatrix} \boldsymbol{0} & \boldsymbol{I} \\ -\boldsymbol{I} & \boldsymbol{0} \end{bmatrix},
\end{equation}
with $\boldsymbol{I}$ and $\boldsymbol{0}$ denoting the $M \times M$ identity and zero matrices, respectively. A multimode Gaussian state is fully characterized by the mean field (first moment), $\bar{\boldsymbol{q}} = \langle \hat{\boldsymbol{q}} \rangle$, and the $2M \times 2M$ covariance matrix (second moment), 
\begin{equation}
	\boldsymbol{V} = {1 \over 2}  \langle \{\hat{\boldsymbol{q}}-\bar{\boldsymbol{q}}, (\hat{\boldsymbol{q}}-\bar{\boldsymbol{q}})^T \}\rangle, 
\end{equation}
which together contain the complete information about the multimode Gaussian state~\cite{Weedbrook12}. Here, $\{ , \}$ denotes the anticommutator. Since $\bar{\boldsymbol{q}}$ can be readily measured by homodyne detection, we focus on characterizing the covariance matrix, which reveals quantum properties such as squeezing~\cite{Roh25, Roslund14, Cai17} and  entanglement~\cite{Simon00,Gerke15}. For physical validity, the covariance matrix must satisfy the uncertainty relation~\cite{Weedbrook12},
\begin{equation}
    \boldsymbol{V}+i \boldsymbol{\Omega} \succeq 0.
    \label{eq:uncertainty}
\end{equation}

Next, we present the decomposition of the covariance matrix $\boldsymbol{V}$. The Williamson decomposition yields
\begin{equation}
    \boldsymbol{V} = \boldsymbol{S} \begin{bmatrix}
    \boldsymbol{\Lambda} & \boldsymbol{0} \\ 
    \boldsymbol{0} & \boldsymbol{\Lambda}
    \end{bmatrix} \boldsymbol{S}^T,
    \label{eq:Williamson}
\end{equation}
where $\boldsymbol{S}$ is a $2M \times 2M$ symplectic matrix (i.e., $\boldsymbol{S}\boldsymbol{\Omega}\boldsymbol{S}^T=\boldsymbol{\Omega}$) and $\boldsymbol{\Lambda} = \mathrm{diag}(\lambda_1,\dots,\lambda_M)$~\cite{Williamson36}. The quantities $\lambda_1,\dots,\lambda_M$ are the symplectic eigenvalues, and the uncertainty relation (Eq.~(\ref{eq:uncertainty})) imposes the constraint $\lambda_m \ge 1$~\cite{Serafini2017}. The Bloch-Messiah decomposition further factorizes the symplectic matrix~\cite{Bloch62} as
\begin{equation}
    \boldsymbol{S} = \boldsymbol{O}_2 \boldsymbol{D}_s \boldsymbol{O}_1,
    \label{eq:BM}
\end{equation}
where $\boldsymbol{D}_s = \mathrm{diag}(e^{r_1},...,e^{r_M}, e^{-r_1},...,e^{-r_M})$ describes a multimode squeezing operation acting independently on each mode, and $\boldsymbol{O}_{1,2}$ are orthogonal symplectic matrices corresponding to linear-optical transformations.

\subsection{Measurements of multimode Gaussian states}

\begin{figure}[bt]%
	\centering
	\includegraphics[width = 85mm]{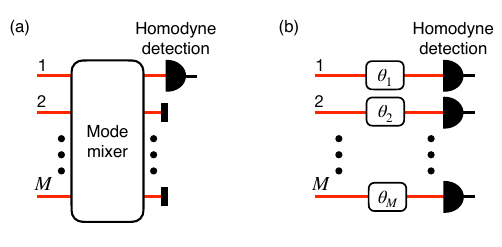}
	\caption{Measurement schemes for characterizing multimode Gaussian states. (a) Single homodyne detection. The quadrature of a selected mode (either an individual mode or a superposition of modes) is measured using a single homodyne detector after a mode mixer.
    (b) Joint homodyne detection. The quadratures of $M$ modes are measured simultaneously using multiple homodyne detectors, each with an adjustable phase $\theta_m$.
    }
	\label{fig:homodyne_detection}
\end{figure}

For the quantum state tomography (QST) of a multimode Gaussian state, we perform informationally-complete quantum measurements. Figure~\ref{fig:homodyne_detection} illustrates two measurement schemes: (a) single homodyne detection and (b) joint homodyne detection. 

In single homodyne detection, a mode mixer provides access to individual modes as well as superpositions of multiple modes. To ensure information completeness, the following quadrature operators are measured:
\begin{equation}
\begin{aligned}
    &\hat{\xi}_{m,n}(\theta) =\begin{cases}
    \hat{q}_m(\theta) & (m=n) \\
    {1\over \sqrt{2}}\hat{q}_m(\theta) + {1\over \sqrt{2}} \hat{q}_n(\theta) & (m < n)
    \end{cases}
    \label{eq:measurement}
\end{aligned}
\end{equation}
for $1 \le m \le n \le 2M$, yielding $N_s=M(2M+1)$ measurement settings. Here, $\hat{q}_m(\theta)$ denotes a quadrature operator with phase shift $\theta$,
\begin{equation}
\hat{q}_m(\theta) = \begin{cases}
\hat{x}_m \cos \theta+\hat{p}_{m} \sin \theta & (m \le M) \\
\hat{p}_{m-M} \cos \theta - \hat{x}_{m-M} \sin \theta &  (m>M),
\end{cases}
\end{equation}
where $\theta$ is either fixed at zero in phase-locked homodyne detection or varied from 0 to $2\pi$ in phase-scanning homodyne detection. $\hat{q}_m(\theta)$ is measured $N_r$ times, resulting in a total of $N_t = N_r N_s$ measurement outcomes. These measurement results are then used to reconstruct the covariance matrix of a multimode Gaussian state~\cite{D'Auria09, Raymer99}.

In joint homodyne detection, quadratures of all $M$ modes are measured simultaneously. Given a measurement setting of $\boldsymbol{\theta} = [\theta_1, ..., \theta_M]$, the measured quadratures are
\begin{equation}
    \hat{\boldsymbol{\zeta}}(\boldsymbol{\theta}) = [\hat{q}_1(\theta_1),...,\hat{q}_M(\theta_M)]^T,
    \label{eq:jointmeasurement}
\end{equation}
resulting in an $M$-dimensional vector operator. For the information completeness, we use the measurement angles of
\begin{equation}
\begin{aligned}
    &\boldsymbol{\theta}^j = \\
    &  \begin{cases}
        [0,...,0] & (j = 1)\\
        {\pi \over 2}[s\left({2 \over {2^{j-1}}}\right),...,s\left({M+1 \over {2^{j-1}}}\right)] & (2 \le j \le \lfloor\log_2 M \rfloor +2) \\
        {\pi \over 4}[1,1,1,...,1] & (j = \lfloor\log_2 M \rfloor +3),
    \end{cases}
\end{aligned}
\label{eq:joint_setting}
\end{equation}
where $s(x) =  2\lfloor x \rfloor - \lfloor2 x \rfloor +1$. The number of measurement settings is $N_s=\lfloor \log_2 M \rfloor +3 $. Table~\ref{tab:measurements} provides an example for $M=8$, listing the measurement settings $\boldsymbol{\theta}^j$ and the corresponding measurement operators. For joint homodyne detection, the total number of measurement outcomes is $N_t = M N_r N_s$, since $M$ quadrature outcomes are simultaneously obtained in each of the $N_s$ measurement settings. These measurement outcomes are used to reconstruct the covariance matrix in the following section.

\begin{table}[t]
\caption{Joint homodyne detection settings for 8 modes.}
\label{tab:measurements}
\begin{tabular}{c|c|c}
\hline
\hline
$j$ & $\boldsymbol{\theta}^j$ & $\hat{\boldsymbol{\zeta}}(\boldsymbol{\theta}^j)$ \\
\hline
1 & $[0,0,0,0,0,0,0,0]$ & $[ \hat{x}_1,\hat{x}_2,\hat{x}_3,\hat{x}_4,\hat{x}_5,\hat{x}_6,\hat{x}_7,\hat{x}_8 ]$\\
\hline
2 & ${\pi \over 2}[0,1,0,1,0,1,0,1]$ & $[ \hat{x}_1,\hat{p}_2,\hat{x}_3,\hat{p}_4,\hat{x}_5,\hat{p}_6,\hat{x}_7,\hat{p}_8 ]$\\
\hline
3 & ${\pi \over 2}[0,0,1,1,0,0,1,1]$ & $[ \hat{x}_1,\hat{x}_2,\hat{p}_3,\hat{p}_4,\hat{x}_5,\hat{x}_6,\hat{p}_7,\hat{p}_8 ]$\\
\hline
4 & ${\pi \over 2}[0,0,0,0,1,1,1,1]$ & $[ \hat{x}_1,\hat{x}_2,\hat{x}_3,\hat{x}_4,\hat{p}_5,\hat{p}_6,\hat{p}_7,\hat{p}_8 ]$\\
\hline
5 & ${\pi \over 2}[1,1,1,1,1,1,1,1]$ & $[ \hat{p}_1,\hat{p}_2,\hat{p}_3,\hat{p}_4,\hat{p}_5,\hat{p}_6,\hat{p}_7,\hat{p}_8 ]$\\
\hline
6 & ${\pi \over 4}[1,1,1,1,1,1,1,1]$ & $[ (\hat{x}_1+\hat{p}_1)/\sqrt{2},..., (\hat{x}_8+\hat{p}_8)/\sqrt{2}]$\\
\hline
\hline
\end{tabular}
\end{table}

\section{Reconstruction Methods}

\subsection{Direct reconstruction}

The conventional approach is to directly compute the second-order moments from the measured quadrature outcomes and reconstruct the covariance matrix~\cite{D'Auria09,Roslund14,Cai17}. This approach is analogous to the linear inversion method used in density-matrix reconstruction~\cite{Schwemmer2015, Aditi2025}. In this section, we provide the detailed procedures of this direct approach for single homodyne detection and joint homodyne detection.

The covariance matrix $\boldsymbol{V}$ of an $M$-mode Gaussian state can be written in block form as
\begin{equation}
    \boldsymbol{V} = \begin{bmatrix}
    \boldsymbol{V}^{xx} & \boldsymbol{V}^{xp} \\
    \boldsymbol{V}^{px} & \boldsymbol{V}^{pp}
    \end{bmatrix},
\end{equation}
where $\boldsymbol{V}^{xx}$ and $\boldsymbol{V}^{pp}$ describe correlations of the $\hat{x}$ and $\hat{p}$ quadratures, respectively, while $\boldsymbol{V}^{xp}$ and $\boldsymbol{V}^{px}$ describe cross-correlations between them.
In single homodyne detection, each block matrix can be obtained from
\begin{widetext}
\begin{equation}
\begin{aligned}
\boldsymbol{V}^{xx}_{m,n} & = \mathrm{Var}(\hat{\xi}_{m,n}(0))-{\mathrm{Var}(\hat{\xi}_{m,m}(0))+\mathrm{Var}(\hat{\xi}_{n,n}(0)) \over 2}(1-\delta_{m,n}) \\
\boldsymbol{V}^{pp}_{m,n} & = \mathrm{Var}(\hat{\xi}_{m+M,n+M}(0))-{\mathrm{Var}(\hat{\xi}_{m+M,m+M}(0))+\mathrm{Var}(\hat{\xi}_{n+M,n+M}(0)) \over 2}(1-\delta_{m,n}) \\
\boldsymbol{V}^{px}_{m,n} & = \mathrm{Var}(\hat{\xi}_{m,n+M}(0))-{\mathrm{Var}(\hat{\xi}_{m,m}(0))+\mathrm{Var}(\hat{\xi}_{n+M,n+M}(0)) \over 2} \\
\boldsymbol{V}_{m,n}^{xp} & = \boldsymbol{V}^{px}_{n,m},
\end{aligned}
\end{equation}
\end{widetext}
where $1 \le m \le M,~~ 1 \le n \le M$, and $\delta_{m,n}$ is the Kronecker delta. Note that the variance $\mathrm{Var}(\hat{\xi}_{m,n}(\theta))$ is directly obtained from the quadrature measurements in Eq.~(\ref{eq:measurement}). In joint homodyne detection, the simultaneously acquired quadrature outcomes enable the reconstruction of $\boldsymbol{V}^{xx}$ and $\boldsymbol{V}^{pp}$. Measurements of the $\hat{x} $ ($\hat{p}$) quadratures for all modes yield their diagonal elements through the variance of each quadrature and off-diagonal elements through the covariance between different modes. For the reconstruction of $\boldsymbol{V}^{xp}$ and $\boldsymbol{V}^{px}$, the off-diagonal elements are similarly obtained from the covariance between $\hat{x}_m$ and $\hat{p}_n$ in different modes ($m\ne n$), while the diagonal elements are determined from the difference between $\mathrm{Var}((\hat{x}_m+\hat{p}_m)/\sqrt{2})$ and $(\mathrm{Var}(\hat{x}_m) + \mathrm{Var}(\hat{p}_m))/2$.

This direct reconstruction method provides a convenient means of reconstructing a covariance matrix from  quadrature outcomes. However, in realistic experiments with a finite number of outcomes, it often yields a unphysical covariance matrix that violates the uncertainty relation~\cite{Mele25, Gerke15, Shchukin24}. Finite sampling induces statistical fluctuations in the symplectic eigenvalues of the reconstructed covariance matrix, and the smallest  eigenvalue may fall below unity. This issue becomes more pronounced as the number of modes increases, since the probability that at least one symplectic eigenvalue falls below unity grows with $M$. Moreover, even when the reconstructed matrix remains physical, the method is not well suited for estimating nonlinear quantities~\cite{Schwemmer2015} such as fidelity~\cite{Banchi15} and entanglement~\cite{Simon00}. A systematic analysis of these limitations is presented in Section~\ref{sec:benchmark}.

\subsection{Maximum-likelihood-estimation reconstruction}

\begin{figure}[tb]%
	\centering
	\includegraphics[width = 82mm]{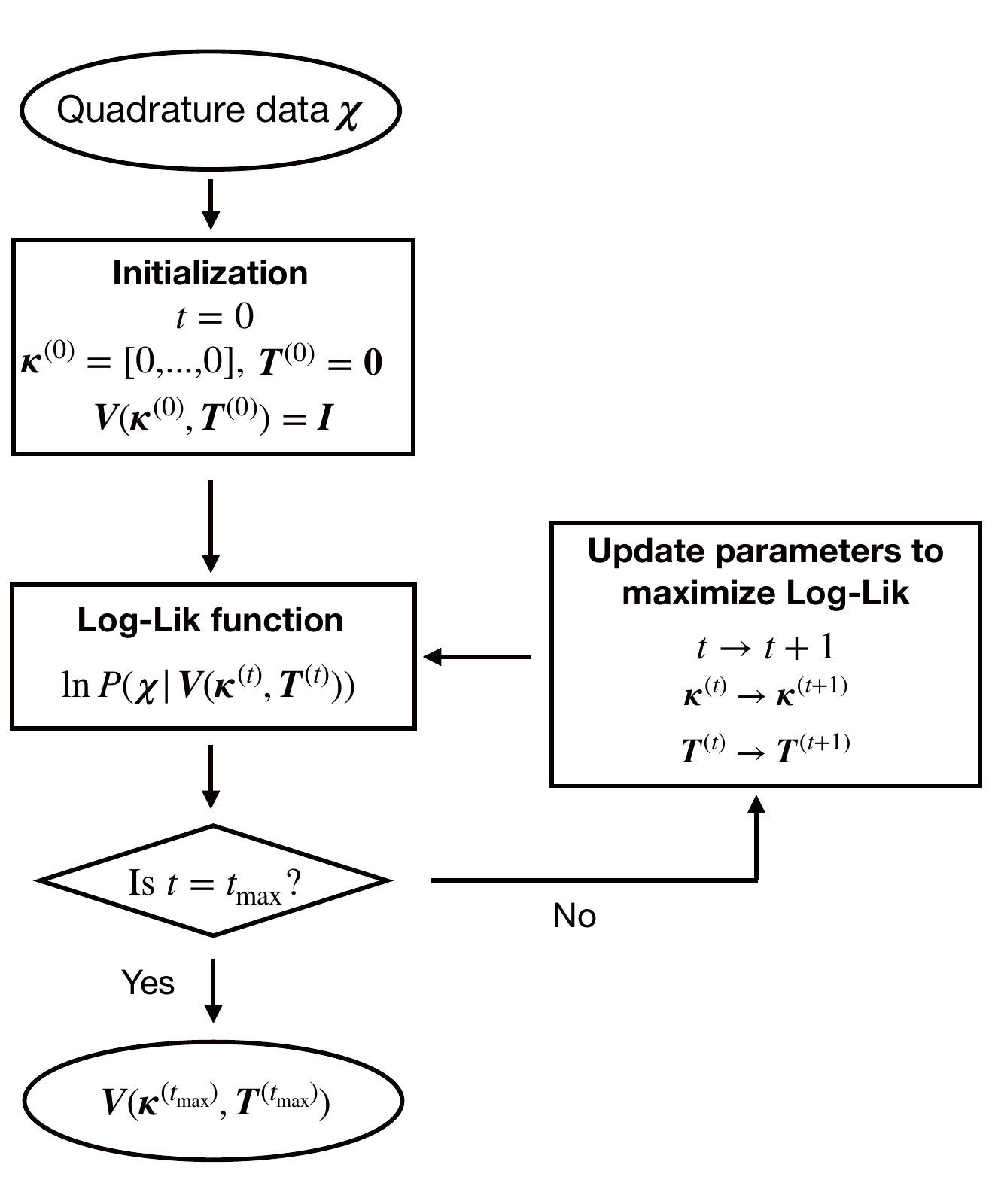} 
	\caption{Illustration of the maximum likelihood estimation method for reconstructing a covariance matrix. A set of quadrature outcomes $\boldsymbol{\chi}$ is obtained using either single homodyne detection or joint homodyne detection (see Fig.~\ref{fig:homodyne_detection}). The covariance matrix is parametrized as $\boldsymbol{V}(\boldsymbol{\kappa}^{(t)}, \boldsymbol{T}^{(t)})$ to express a general multimode Gaussian state while satisfying the physical condition. The likelihood function is computed in each step $t$ based on the parameters $\boldsymbol{\kappa}^{(t)},\boldsymbol{T}^{(t)}$ and the quadrature outcomes $\boldsymbol{\chi}$. The parameters are updated to maximize the likelihood function. After all iterations, the final parameters $\boldsymbol{\kappa}^{(t_\mathrm{max})}$ and $\boldsymbol{T}^{(t_\mathrm{max})}$ are used to reconstruct the covariance matrix $\boldsymbol{V}(\boldsymbol{\kappa}^{(t_\mathrm{max})}, \boldsymbol{T}^{(t_\mathrm{max})})$.
	}
	\label{fig:scheme}
\end{figure}

To address unphysical covariance-matrix reconstructions in quantum state tomography, we employ the MLE method~\cite{James01,Hradil1997}. The central idea is to parametrize the covariance matrix so that it satisfies the uncertainty relation by construction and to determine the parameters that maximize the likelihood of the observed quadrature outcomes. We build on the single homodyne detection method in Ref.~\cite{Roh25} and extend it to the joint homodyne detection scheme.

Figure~\ref{fig:scheme} illustrates the overall process. A set of quadrature outcomes $\boldsymbol{\chi}$ is obtained via single or joint homodyne detection. For the parametrization, we adopt the Williamson decomposition in Eq.~(\ref{eq:Williamson}). The physical constraint on the symplectic eigenvalues, $\lambda_m \ge 1$, is ensured by parametrizing them as $\lambda_m = \kappa_m^2+1$ with $\kappa_m \in \mathbb{R}$. The $2M \times 2M$ symplectic matrix $\boldsymbol{S}$ is parametrized by a real and symmetric matrix $\boldsymbol{T}$ ~\cite{Petrov20},
\begin{equation}
    \boldsymbol{S}=\left(\boldsymbol{I}+{\boldsymbol{\Omega} \boldsymbol{T} \over 2}\right) \left(\boldsymbol{I}-{\boldsymbol{\Omega} \boldsymbol{T} \over 2}\right)^{-1}.
    \label{eq:sym_para}
\end{equation}
In this way, any $2M \times 2M$ covariance matrix $\boldsymbol{V}$ can be parametrized by an $M$-dimensional vector $\boldsymbol{\kappa} = [\kappa_1,...,\kappa_M]$ and the $\boldsymbol{T}$ matrix containing $M(2M+1)$ parameters. Accordingly, the total number of parameters is $2M(M+1)$.

For reconstructing the covariance matrix, the parameters, $\boldsymbol{\kappa}$ and $\boldsymbol{T}$, are updated to maximize the likelihood of obtaining the outcome set $\boldsymbol{\chi}$ under the given measurement scheme. Specifically, the likelihood function at the iteration step $t$ is given by $P(\boldsymbol{\chi} \vert \boldsymbol{V}^{(t)})$, where $\boldsymbol{V}^{(t)}=\boldsymbol{V} (\boldsymbol{\kappa}^{(t)},\boldsymbol{T}^{(t)})$ is the parametrized covariance matrix. As the initial parameters, we choose $\boldsymbol{\kappa}^{(0)} = [0,...,0]$ and $\boldsymbol{T}^{(0)} = \boldsymbol{0}$, corresponding to the vacuum covariance matrix $\boldsymbol{V}^{(0)} = \boldsymbol{I}$.

In single homodyne detection, the set of quadrature outcomes is denoted by $\boldsymbol{\chi} = \{\xi_{m,n}(\theta_k)\}$,
where $\xi_{m,n}(\theta_k)$ represents the measurement outcome of $\hat{\xi}_{m,n}(\theta_k)$ defined in Eq.~(\ref{eq:measurement}), and $k = 1,\ldots,N_r$ indexes the repeated measurements. The log-likelihood function is given by
\begin{equation}
    \ln P(\boldsymbol{\chi} \vert \boldsymbol{V}^{(t)} ) =
    \sum_{m=1}^{2M} \sum_{n=m}^{2M} \sum_{k=1}^{N_r} \ln \left[ {\mathrm{exp} \left(-{\xi^2_{m,n} (\theta_k) \over 2 \sigma_{m,n}^2(\theta_k)}\right) \over \sqrt{2 \pi} \sigma_{m,n}(\theta_k)} \right],
    \label{eq:loglik-single}
\end{equation}
where $\sigma_{m,n}^2 (\theta_k)$ denotes the variance of $\hat{\xi}_{m,n} (\theta_k)$ evaluated for $\boldsymbol{V}^{(t)}$. In joint homodyne detection, the set of quadrature outcomes is $\boldsymbol{\chi}  = \{\boldsymbol{\zeta}_k(\boldsymbol{\theta}^j)\}$, where $\boldsymbol{\zeta}_k(\boldsymbol{\theta}^j)$ denotes the $M$-dimensional vector of outcomes obtained by measuring $\hat{\boldsymbol{\zeta}}(\boldsymbol{\theta}^j)$ in Eq.~(\ref{eq:jointmeasurement}). The phase setting $j$ is specified in Eq.~(\ref{eq:joint_setting}), and $k=1,\ldots,N_r$ indexes the repeated measurements. The log-likelihood function is
\begin{equation}
    \begin{split}
	& \ln P(\boldsymbol{\chi} \vert \boldsymbol{V}^{(t)} ) = \\ 
    & \sum_{j=1}^{N_s} \sum_{k=1}^{N_r}
	\ln \left[{
	\exp \left(-{1 \over 2} \boldsymbol{\zeta}^T_k(\boldsymbol{\theta}^j) ~(\boldsymbol{\Sigma}^{(t)}_{\boldsymbol{\theta}^j})^{-1} ~ \boldsymbol{\zeta}_k (\boldsymbol{\theta}^j) \right)
	\over 
	(2\pi)^{M/2}~(\mathrm{det}\boldsymbol{\Sigma}^{(t)}_{\boldsymbol{\theta}^j} )^{1/2}
		}\right].
	\label{eq:loglik2}
    \end{split}
\end{equation}
Here $\boldsymbol{\Sigma}^{(t)}_{\boldsymbol{\theta}^j}$ is an $M \times M$ reduced covariance matrix for the quadrature measurement $\hat{\boldsymbol{\zeta}}(\boldsymbol{\theta}^j)$, given by
\begin{equation}
	\boldsymbol{\Sigma}^{(t)}_{\boldsymbol{\theta}^j}=
	\begin{bmatrix}
	\cos \boldsymbol{\theta}^j & \sin \boldsymbol{\theta}^j
	\end{bmatrix} ~\boldsymbol{V}^{(t)}
	\begin{bmatrix}
	\cos \boldsymbol{\theta}^j \\  \sin \boldsymbol{\theta}^j
	\end{bmatrix},
\end{equation}
where $\cos \boldsymbol{\theta} = \mathrm{diag}( \cos \theta_1,\dots, \cos \theta_M)$, $\sin \boldsymbol{\theta} = \mathrm{diag}(\sin \theta_1,\dots, \sin \theta_M)$.

To update the parameters $\boldsymbol{\kappa}$ and $\boldsymbol{T}$, we perform gradient ascent using the Adam optimizer~\cite{Kingma14}. Compared with a basic gradient ascent method with a fixed step size, the Adam optimizer generally improves convergence efficiency by incorporating momentum and adaptive step-size adjustment. Let $u$ denote each parameter in $\boldsymbol{\kappa}$ and $\boldsymbol{T}$. The parameter $u$ is updated as follows:
\begin{equation}
    u^{(t)} = u^{(t-1)}+\eta {m^{(t)}/(1-(\beta_1)^t) \over \sqrt{v^{(t)}/(1-(\beta_2)^t)+\epsilon}}.
\end{equation}
Here, $t$ is the iteration step, and $m^{(t)}$ ($v^{(t)}$) is the first (second) moment of the Adam optimizer at the $t$-th step, which is defined as below recurrence formula
\begin{equation}
    \begin{aligned}
    m^{(t)} & = \beta_1 m^{(t-1)} + (1-\beta_1) {\partial \ln P(\boldsymbol{\chi} \vert \boldsymbol{V}^{(t-1)}) \over \partial u} \bigg|_{u^{(t-1)}} \\
    v^{(t)} & = \beta_2 v^{(t-1)} + (1-\beta_2) \left({\partial \ln P(\boldsymbol{\chi} \vert \boldsymbol{V}^{(t-1)}) \over \partial u} \bigg|_{u^{(t-1)}}\right)^2.
    \end{aligned}
\end{equation}
. The first and second moments are initialized to zero ($m^{(0)} = v^{(0)}=0$). We adopt the commonly used hyperparameters $\beta_1 = 0.9$, $\beta_2 = 0.999$, and $\epsilon = 10^{-8}$~\cite{Kingma14}. We use the learning rate of $\eta=0.02$, which yields reasonably fast convergence without noticeable oscillations near convergence. The total number of iterations ($t_\mathrm{max}$) is set to 500 for single homodyne detection and 200 for joint homodyne detection. We will discuss the reliability of this method in Section \ref{sec:benchmark}.

\begin{figure}[b]%
	\centering
	\includegraphics[width = 85mm]{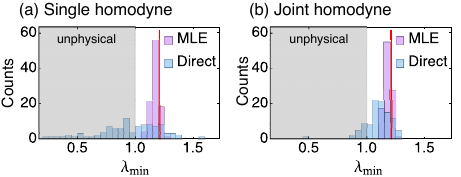}
	\caption{Minimum symplectic eigenvalues ($\lambda_{\min}$) of reconstructed covariance matrices obtained with (a) single homodyne detection and (b) joint homodyne detection. A two-mode cluster state with 6-dB pure squeezing and 30\% optical loss is considered. Histograms of $\lambda_{\min}$ are obtained from 100 Monte Carlo simulations, each generating a total of $N_t=10,000$ measurement outcomes. The same simulation data are used for the direct method (blue bars) and the MLE method (purple bars). Gray regions indicate unphysical reconstructions ($\lambda_{\min}<1$), and red lines indicate the true value of $\lambda_{\min}$.}
	\label{fig:physical}
\end{figure}

\begin{figure*}[tb]%
	\centering
	\includegraphics[width = 170mm]{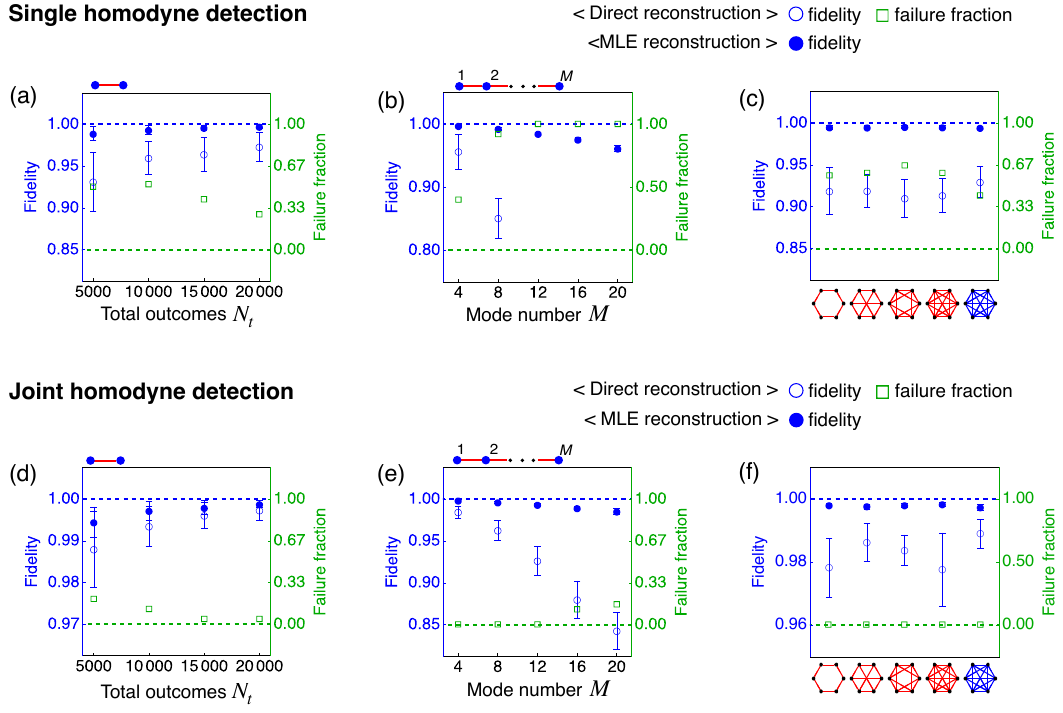} 
	\caption{
    Performance of covariance-matrix reconstruction methods. (a-c) single and (d-f) joint homodyne detection. Open and filled markers denote the direct and MLE reconstruction methods, respectively, and blue circles and green squares represent the fidelity (with respect to the true state) and the fraction of failures ($\lambda_{\min}<1$), respectively. The MLE method does not produce any failure cases.
    (a,d) Reconstruction of a two-mode cluster state as a function of the total number of outcomes $N_t$.
    (b,e) Reconstruction of linear cluster states as the mode number $M$ increases. The number of repeated measurements $N_r$ is fixed at 5,000, and the total number of outcomes is $N_t = N_r M(2M+1)$ for single homodyne detection and $N_t = N_r M (\lfloor\log_2 M \rfloor +3)$ for joint homodyne detection.
    (c,f) Reconstruction of six-mode entangled states with various connectivities. The x-axis labels indicate different correlation types: red graphs for graph states and the blue graph for a GHZ state. The number of repeated measurements is $N_r=$5,000. In all simulations, the quantum states are constructed from 6-dB pure squeezing and 30\% optical loss (Appendix~A). Error bars and failure fractions are obtained from repeated simulations: 50 runs for (a,c,d,f) and 25 runs for (b,e).
    }
	\label{fig:performance}
\end{figure*}

\section{Reconstruction Performance} \label{sec:benchmark}

We analyze the performance of the direct and the MLE reconstruction methods. For this purpose, Monte Carlo simulations are performed for various Gaussian states under the measurement schemes shown in Fig.~\ref{fig:homodyne_detection}. The detailed procedures for constructing covariance matrices used in the simulations are described in Appendix~A. For a fair comparison, both reconstruction methods are applied to the same simulation data.

First, we verify the physical validity of the reconstructed covariance matrices based on the uncertainty relation in Eq.~(\ref{eq:uncertainty}), i.e., $\lambda_m \ge 1$. Specifically, we compute the minimum symplectic eigenvalue ($\lambda_{\min}$) of each reconstructed covariance matrix. Figure~\ref{fig:physical} shows the results of 100 repeated simulations for a two-mode cluster state under (a) single homodyne detection and (b) joint homodyne detection. For both detection schemes, the total number of measurement outcomes is fixed at $N_t=$ 10,000. In the direct reconstruction, 53 and 17 out of 100 simulation results violate the uncertainty relation ($\lambda_{\min} < 1$) for single and joint homodyne detection, respectively. In contrast, the MLE reconstruction always yields a physical covariance matrix. Moreover, the minimum symplectic eigenvalues lie close to the true values (red lines), whereas those obtained from the direct reconstruction exhibit a broader and biased distribution.

We next investigate how the fidelity of the reconstructed covariance matrix varies with the total number of outcomes $N_t$. Throughout this work, we employ the Uhlmann fidelity (the square of the Bures fidelity) for the fidelity calculation ~\cite{Banchi15}.
Figure~\ref{fig:performance}(a) and (d) show the results for the two-mode cluster state using single and joint homodyne detection, respectively. As already observed in Fig~\ref{fig:physical}, the direct reconstruction often yields unphysical covariance matrices, which are therefore excluded from the fidelity calculation. Clearly, the MLE reconstruction yields higher fidelity than the direct method for both measurement schemes. In the direct reconstruction, as $N_t$ decreases, the fidelity decreases and the fraction of unphysical reconstructions increases. In contrast, the MLE method maintains high fidelity even for small $N_t$. Finally, joint homodyne detection outperforms single homodyne detection because the simultaneously obtained quadrature outcomes provide additional correlation information between different quadratures.

We further study the reconstruction performance as the mode number $M$ increases. Linear cluster states with 4, 8, 12, 16, and 20 modes are considered, as shown in Fig.~\ref{fig:performance}(b) for single homodyne detection and Fig.~\ref{fig:performance}(e) for joint homodyne detection. In the direct reconstruction, the failure fraction increases and the fidelity rapidly decreases as $M$ increases. In particular, under single homodyne detection, the direct method never produces a physical covariance matrix for $M \ge 12$. Under joint homodyne detection, the direct method begins to produce unphysical covariance matrices from 16 modes. In contrast, the MLE method always reconstructs physical covariance matrices and maintains high fidelities, outperforming the direct method.

Entangled states of different connectivities are examined in Fig.~\ref{fig:performance}(c,f). We consider six-mode graph states~\cite{vanLoock2007} and a six-mode GHZ state~\cite{Loock00}. In the direct reconstruction, single homodyne detection frequently yields unphysical results and low fidelities, whereas joint homodyne detection achieves higher fidelity without producing unphysical results. The performance is substantially improved in the MLE reconstruction: both single and joint homodyne detection always produce physical covariance matrices while maintaining very high fidelity.

\begin{figure}[t]%
	\centering
	\includegraphics[width = 85 mm]{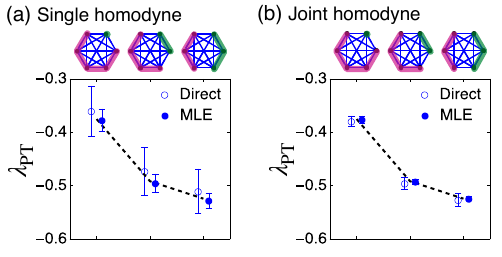}
	\caption{Entanglement detection based on covariance-matrix reconstruction. The covariance matrices of the GHZ state from Fig.~\ref{fig:performance}(c,f) are used for (a) single and (b) joint homodyne detection, where the unphysical covariance matrices are excluded in the calculation. The minimum eigenvalue under partial transposition ($\lambda_{\textrm{PT}}$) is plotted for the bipartition represented by the pink and green shaded areas in each graph diagram. Open and filled circles correspond to the results obtained with the direct and MLE methods, respectively. Black dashed lines indicate the true values.
    }
	\label{fig:entanglement}
\end{figure}

Finally, we investigate the performance of the reconstruction methods for entanglement detection (Fig.~\ref{fig:entanglement}). The bipartite entanglement criterion based on partial transposition is applied to the reconstructed covariance matrices of the six-mode GHZ state~\cite{Simon00}. In all cases, negative eigenvalues occur under partial transposition, successfully detecting bipartite entanglement. In particular, the MLE method yields more reliable results than the direct method, showing better agreement with the true values and exhibiting smaller errors.


\section{Experimental quantum state tomography}  \label{sec:experiment}

\begin{figure}[b]%
	\centering
	\includegraphics[width = 90mm]{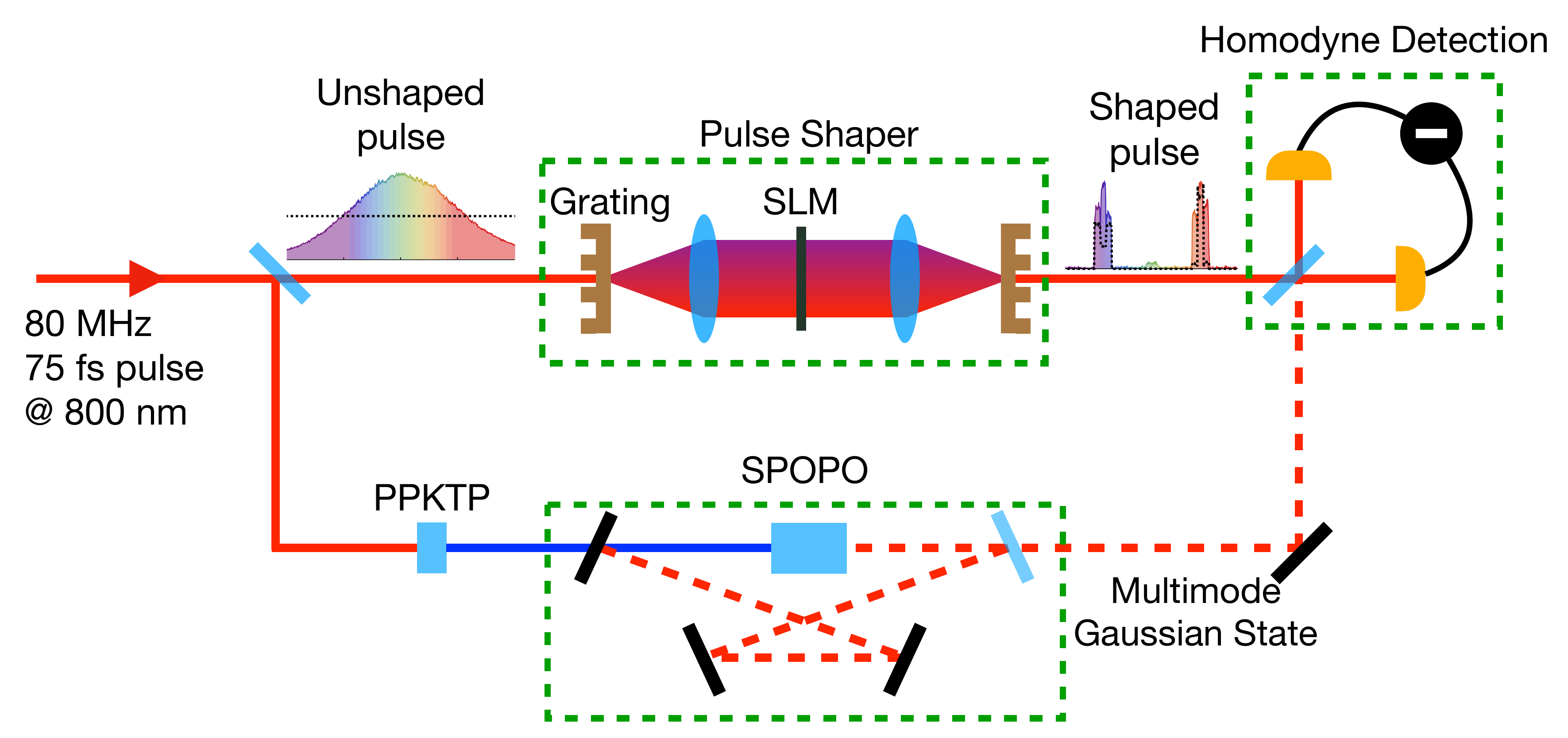}
	\caption{Experimental setup. A synchronously pumped optical parametric oscillator (SPOPO) generates multimode quantum light in time-frequency modes. A pulse shaper prepares a local oscillator in a complex time-frequency mode. Homodyne detection with the local oscillator enables quadrature measurements $\hat{\xi}_{m,n}(\theta)$ in Eq.~(\ref{eq:measurement}), i.e., single homodyne detection, providing complete information for covariance-matrix reconstruction.
    }
	\label{fig:exp_setup}
\end{figure}

We now apply the reconstruction methods to characterize multimode Gaussian states generated experimentally. The various six-mode entangled states shown in Fig.~\ref{fig:performance}(c,f) are  generated and characterized experimentally, and the performance of the direct and MLE reconstruction methods is compared. Furthermore, we extend the experiment to a fully connected ten-mode graph state and introduce an analysis method to identify the multimode structure of squeezing and thermal noise.

Figure~\ref{fig:exp_setup} illustrates the experimental setup used to generate and characterize multimode quantum states. We investigate multimode quantum light in time-frequency modes, where access to complex mode bases enables the realization of various entangled states~\cite{Cai17,Roh25}. A synchronously pumped optical parametric oscillator (SPOPO) generates such multimode quantum light, and homodyne detection with a pulse shaper measures the quantum light in complex mode bases. A detailed description of the experimental setup is provided below.

The fundamental laser source is a Ti:sapphire femtosecond laser that generates a train of femtosecond pulses (duration: 75 fs, central wavelength: 800 nm) with a repetition rate of 80 MHz. The pump laser for the SPOPO is prepared by second-harmonic generation of the fundamental laser using a 1-mm periodically poled potassium titanyl phosphate (PPKTP) crystal. The SPOPO is a ring-type cavity containing a 4-mm-thick PPKTP crystal for type-0 phase matching, and the cavity length is locked to the Ti:sapphire laser cavity using the Pound-Drever-Hall technique. The SPOPO generates multimode quantum light in time-frequency modes.

To address the quantum light in complex mode bases, we employ homodyne detection that measures a quadrature operator in a desired time-frequency mode. Since the measurement mode is determined by the local oscillator (LO), we construct a pulse shaper to control the time-frequency mode of the LO. The pulse shaper has a folded 4$f$-configuration consisting of an optical diffraction grating, a cylindrical lens, and a spatial light modulator. The spectral resolution of the pulse shaper is 0.07 nm, with a control range of 795--805 nm. In the homodyne detection, the mode-matching visibility is 95 \%, the quantum efficiency of the photodiodes is 99 \%, and the electronic circuit bandwidth is 12 MHz. Quadrature outcomes are obtained at 1.4 MHz sidebands with a bandwidth of 50 kHz. The homodyne phase $\theta$ is obtained by alternating the LO between a reference mode (a Gaussian spectrum centered at 800 nm with a full width at half maximum of 2 nm) for phase estimation and a measurement mode for homodyne detection, each maintained for 80 ms.

\begin{figure*}[tp]
	\centering
	\includegraphics[width = 140mm]{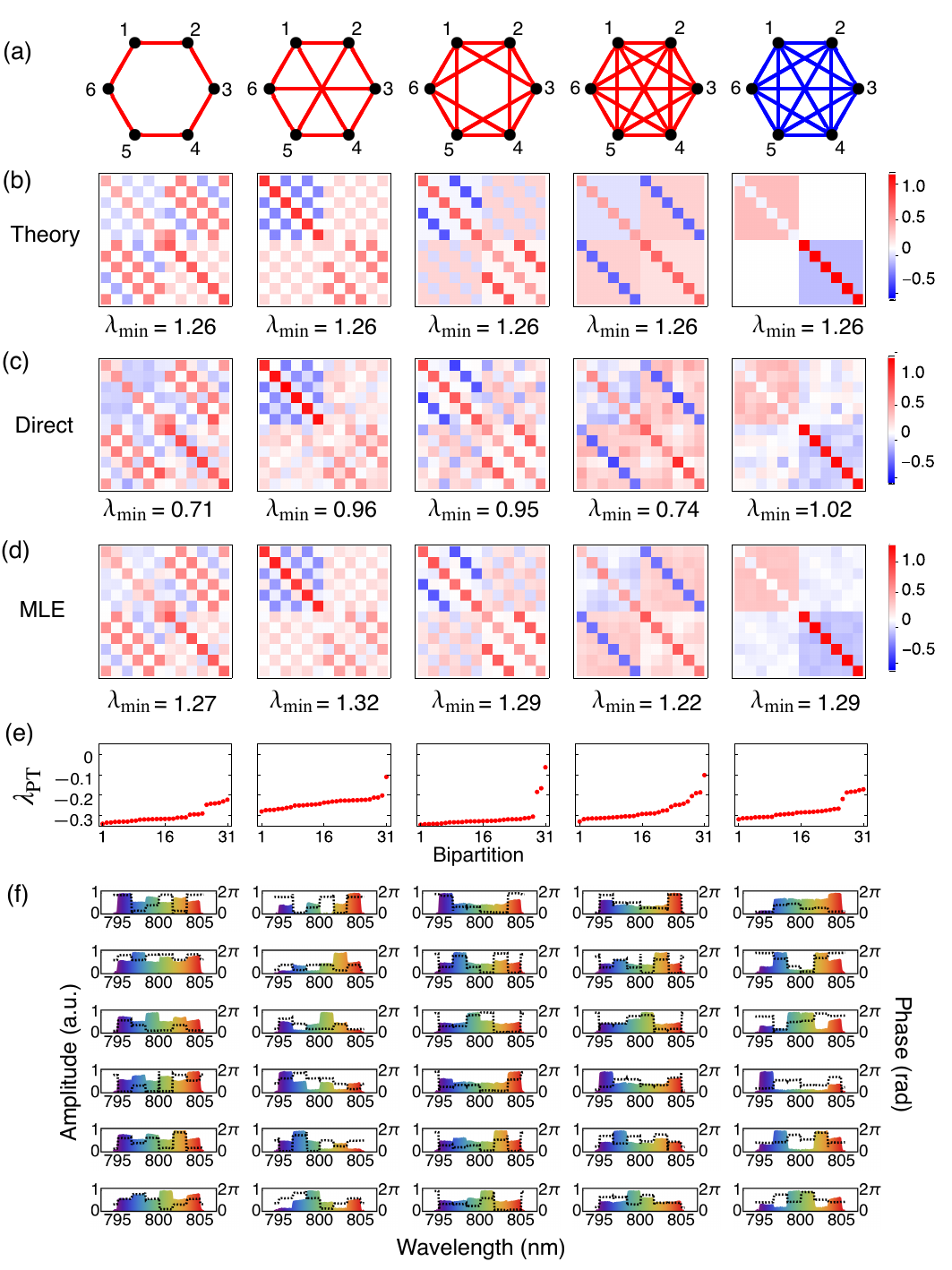} 
	\caption{Experimental quantum state tomography. (a) Graph representation of six-mode graph states (red graphs) and the six-mode GHZ state (blue graph). (b) Theoretical covariance matrices, constructed from 6.1-dB pure squeezing and 51\% optical loss. (c,d) Experimental covariance matrices reconstructed using the direct method (c) and the MLE method (d). Quadrature outcomes are obtained by single homodyne detection, and the total number of measurement outcomes is $N_t = N_rN_s=$ 1,248,000, where $N_r=$16,000 and $N_s=$ 78. $\lambda_{\min}$ is the minimum symplectic eigenvalue of a covariance matrix, requiring $\lambda_{\min} \ge 1 $ for physical validity.
    The fidelities of the physical covariance matrices in (d) with respect to the theoretical ones are 0.97, 0.96, 0.98, 0.98, and 0.97, respectively. For clarity of presentation, the identity matrix is subtracted from all covariance matrices. (e) Entanglement detection based on (d). The minimum eigenvalue under partial transposition is calculated for all possible bipartitions. (f) Time-frequency mode bases used for the experiments. Colored plots and black dashed lines indicate spectral amplitude and phase, respectively. The average mode crosstalks are 0.2\%, 0.3\%, 0.1\%, 4.1\%, and 0.4\%, respectively.
    }
	\label{fig:exp_6}
\end{figure*}

Figure~\ref{fig:exp_6} shows experimental tomography results of six-mode entangled states with various connectivities. To access the quantum states, we measure the quadrature operators $\hat{\xi}_{m,n}(\theta)$ in Eq.~(\ref{eq:measurement}) in the time-frequency mode bases shown in Fig.~\ref{fig:exp_6}(f). In Fig.~\ref{fig:exp_6}(b), the theoretical covariance matrices depend on the connectivity and the type of entanglement. The theoretical covariance matrices satisfy the physical condition, $\lambda_{\min}=1.26 \ge 1$. The experimental covariance matrices reconstructed by the direct method are presented in Fig.~\ref{fig:exp_6}(c). The reconstructed covariance matrices exhibit statistical errors in their elements, deviating from the theoretical ones. Moreover, except for the GHZ state, the covariance matrices do not satisfy the physical condition, making them unsuitable for further analysis. The MLE method resolves the issue. The reconstructed covariance matrices shown in Fig.~\ref{fig:exp_6}(d) all satisfy the physical condition, exhibiting values of $\lambda_{\min}$ similar to the theoretical ones. Moreover, the covariance matrix elements become more stable and agree well with the theoretical ones. This improvement is confirmed by the high fidelities of the reconstructed covariance matrices with respect to the theoretical ones. The generated states exhibit entanglement for all bipartitions, with negative eigenvalues under partial transposition shown in Fig.~\ref{fig:exp_6}(e).

\begin{figure*}[tp]%
	\centering
	\includegraphics[width = 140mm]{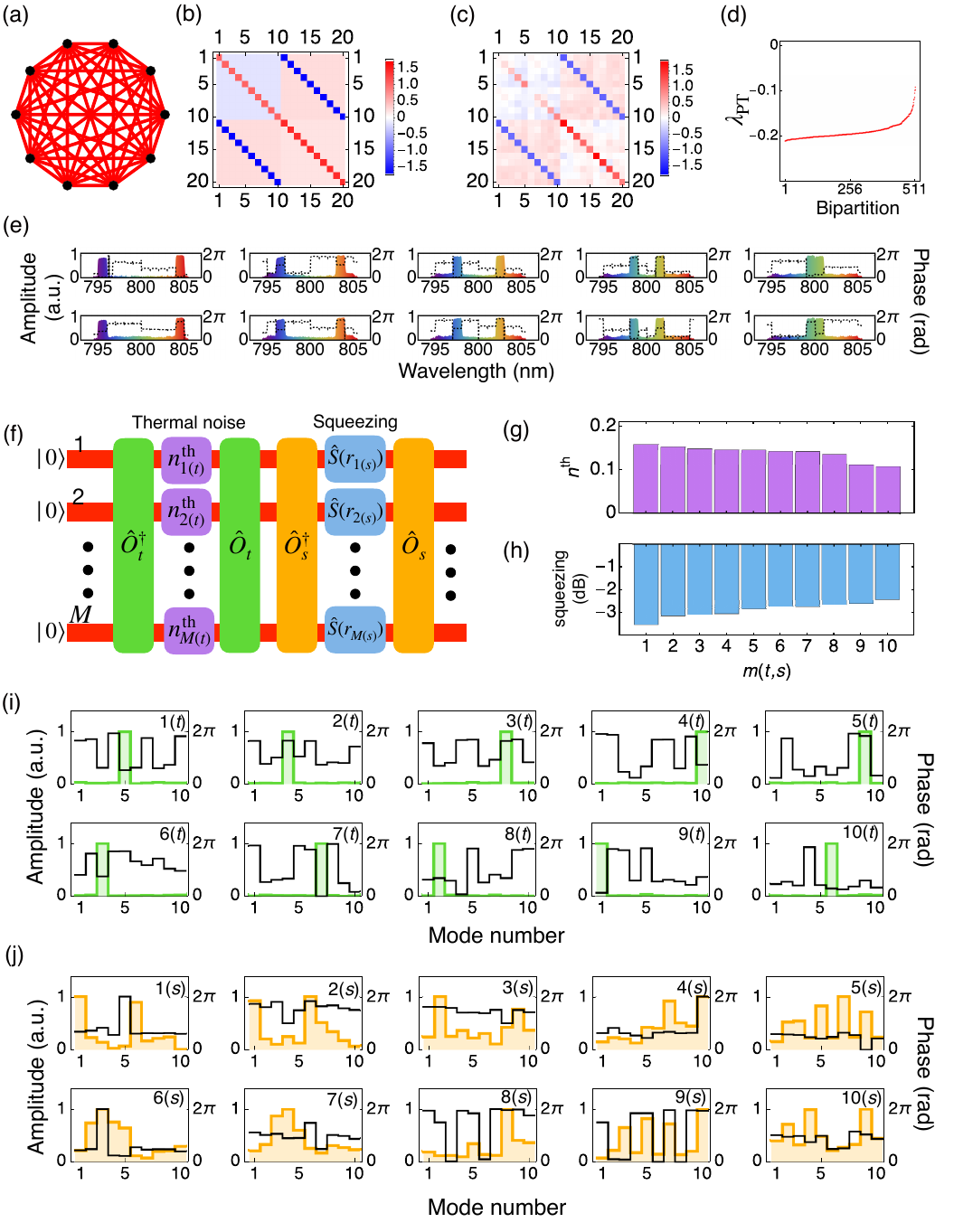} 
	\caption{
    Quantum state tomography of a fully connected ten-mode graph state. (a) Graph representation. (b) Theoretical covariance matrix, constructed from 6.5-dB pure squeezing and 62\% optical loss. (c) Experimentally reconstructed covariance matrix using the MLE method. The identity matrix is subtracted for clarity. The fidelity with respect to the theoretical matrix is 0.87. (d) Entanglement detection based on the experimentally reconstructed covariance matrix. The minimum eigenvalue under partial transposition is calculated for all possible bipartitions. (e) Time-frequency mode basis of the graph state. Colored plots and black dashed lines indicate spectral amplitude and phase, respectively. The average mode crosstalk is as small as 1\%.
    (f) Decomposition of a multimode Gaussian state. Subscripts $t$ and $s$ denote the thermal-noise mode and the squeezing mode, respectively. $n^{\mathrm{th}}_{m(t)}$ is mean photon number in the $m(t)$-th thermal-noise mode, and $r_{m(s)}$ is the squeezing parameter of $m(s)$-th squeezing mode. (g) Mean photon numbers in the thermal-noise modes. (h) Squeezing levels in the squeezing modes. (i) Thermal-noise modes. (j) Squeezing modes.
    }
	\label{fig:exp_10}
\end{figure*}

Having confirmed the advantages of the MLE reconstruction in quantum state tomography, we next apply the method to characterize a fully connected ten-mode graph state. Figure~\ref{fig:exp_10}(a) shows the graph structure, while the corresponding theoretical covariance matrix is shown in Fig.~\ref{fig:exp_10}(b). The experimentally obtained covariance matrix in Fig.~\ref{fig:exp_10}(c) exhibits the expected correlations of a fully connected graph state, with a high fidelity of 0.87 with respect to the theoretical one. It satisfies the physical condition, $\lambda_{\min}=1.21 \ge 1$, and exhibits negative eigenvalues under partial transposition for all bipartitions (Fig.~\ref{fig:exp_10}(d)). The time-frequency mode basis for the ten-mode graph state is illustrated in Fig.~\ref{fig:exp_10}(e).

We further analyze the covariance matrix to characterize the multimode structure of the generated state. For this purpose, we start from the Williamson decomposition in Eq.~(\ref{eq:Williamson}) and the Bloch-Messiah decomposition in Eq.~(\ref{eq:BM}) to express the covariance matrix as
\begin{equation}
    \boldsymbol{V} = \boldsymbol{O}_2 \boldsymbol{D}_s \boldsymbol{O}_1 \begin{bmatrix} \boldsymbol{\Lambda} & \boldsymbol{0} \\ \boldsymbol{0} & \boldsymbol{\Lambda}\end{bmatrix} \boldsymbol{O}_1^T \boldsymbol{D}_s \boldsymbol{O}_2^T.
    \label{eq:WBM}
\end{equation}
This conventional form allows us to interpret the covariance matrix as arising from thermal noise ($\boldsymbol{\Lambda}$), followed by a linear-optical transformation $\boldsymbol{O}_1$, a separable squeezing operation $\boldsymbol{D}_s$, and another linear-optical transformation $\boldsymbol{O}_2$. On the other hand, the following alternative expression directly reveals the multimode structure of the state:
\begin{equation}
    \boldsymbol{V} = \left(\boldsymbol{O}_s \boldsymbol{D}_s \boldsymbol{O}_s^{T} \right) \left( \boldsymbol{O}_t \begin{bmatrix} \boldsymbol{\Lambda} & \boldsymbol{0} \\ \boldsymbol{0} & \boldsymbol{\Lambda}\end{bmatrix} \boldsymbol{O}_t^T \right) \left( \boldsymbol{O}_s \boldsymbol{D}_s \boldsymbol{O}_s^T \right),
    \label{eq:decomposition}
\end{equation}
where we define $\boldsymbol{O}_s = \boldsymbol{O}_2$ and $\boldsymbol{O}_t = \boldsymbol{O}_2 \boldsymbol{O}_1$. Accordingly, the covariance matrix $\boldsymbol{V}$ can be interpreted as thermal noise $\boldsymbol{\Lambda}$ in the mode basis defined by $\boldsymbol{O}_t$, followed by a separable squeezing operation in the mode basis defined by $\boldsymbol{O}_s$. Figure~\ref{fig:exp_10}(f) illustrates this decomposition. The mode basis defined by an orthogonal symplectic matrix $\boldsymbol{O}$ is given by the column vectors of $\boldsymbol{X}-i\boldsymbol{Y}$ in the following representation:
\begin{equation}
    \boldsymbol{O} = \begin{bmatrix}
    \boldsymbol{X} & -\boldsymbol{Y} \\
    \boldsymbol{Y} & \boldsymbol{X}
    \end{bmatrix}.
\end{equation}
Figure~\ref{fig:exp_10}(g--j) shows the results of this multimode analysis. For the thermal noise, the mean photon numbers ($n^\mathrm{th}_m = (\lambda_m-1)/2$) are plotted in (g), with the corresponding noise modes shown in (i). For the multimode squeezing, the squeezing levels are presented in decibels ($10 \log_{10} e^{-2r_m}$) in (h), with the corresponding squeezing modes shown in (j). The spectra of the thermal photon number and the squeezing level are nearly uniform, with a slight decreasing trend over their respective mode bases. The thermal modes exhibit negligible intermodal coupling, whereas the squeezing modes displays intermodal couplings to realize a fully connected graph state.

\section{Conclusion}

We have presented an MLE-based tomography method for the full characterization of multimode Gaussian states under two measurement schemes: single and joint homodyne detection. An efficient parametrization of general multimode Gaussian states renders the MLE approach highly effective for multimode tomography. A systematic analysis demonstrates that the MLE method outperforms the conventional direct reconstruction method, consistently yielding physical states and achieving closer agreement with the true states, as quantified by the fidelity, the minimum symplectic eigenvalue, and the minimum eigenvalue under partial transposition. Between the two measurement schemes, joint homodyne detection exhibits superior performance, benefiting from the correlation information contained in simultaneously obtained quadrature outcomes. The MLE method is further applied to experimentally characterize multimode Gaussian states, including graph states with different connectivity and a GHZ state. The experimental results enable the identification of covariance matrices, entanglement properties, and the multimode structure of squeezing and noise. The availability of the full covariance matrix provides a basis for further analysis, including the investigation of entropy~\cite{Serafini2017}, genuine multipartite entanglement~\cite{Loock15}, quantum steering~\cite{Kogias15}, and quantum Fisher information~\cite{Jiang14}.

By addressing the scalability challenge of multimode quantum tomography~\cite{Gebhart23}, our results suggest a promising direction for completely characterizing large-scale quantum systems. Given an efficient parametrization of other quantum systems~\cite{Ra:2020gg,Gwak25,Ra:2017ia}, the MLE approach is expected to provide a similarly reliable and scalable framework for multimode quantum tomography~\cite{Mele25, Anshu2023}. This technique can play an important role in the development of quantum technologies at scale.

\section*{Acknowledgements}
This work was supported by the Ministry of Science and ICT (MSIT) of Korea (RS-2025-00562372, RS-2024-00408271, RS-2024-00442762, RS-2023-NR119925) under the Information Technology Research Center (ITRC) support program (IITP-2026-RS-2020-II201606), Institute of Information \& Communications Technology Planning \& Evaluation (IITP) grant (RS-2025-25464959, RS-2022-II221029). C. R. acknowledges support from the KAIST Jang Young Sil Fellow Program.
\\
\\

\section*{Appendix A: Constructing covariance matrices} \label{appendix:A}
We provide a detailed method for constructing covariance matrices in Section~\ref{sec:benchmark} (for the Monte Carlo simulations) and in Section~\ref{sec:experiment} (for comparison with the experimental results). We start from the covariance matrix representation in Eq. (\ref{eq:decomposition}), and set the symplectic eigenvalues to one (i.e., $\boldsymbol{\Lambda}=\boldsymbol{I}$). The covariance matrix is then given by
\begin{equation}
		\boldsymbol{V} = \boldsymbol{O}_s \boldsymbol{D}_s^2 \boldsymbol{O}_s^T ,
        \label{eq:pureV}
\end{equation}
where $\boldsymbol{D}_s^2 = \mathrm{diag}(e^{2r},...,e^{2r}, e^{-2r},...,e^{-2r})$ is the covariance matrix of squeezed vacua. $\boldsymbol{O}_s$ determines the type and connectivity of a multimode Gaussian state. For a graph state with an adjacency matrix $\boldsymbol{G}$,
\begin{equation}
	\boldsymbol{O}_s = \begin{bmatrix}
	\boldsymbol{X} & -\boldsymbol{Y} \\ 
	\boldsymbol{Y} & \boldsymbol{X}
	\end{bmatrix},
\end{equation}
where $\boldsymbol{X} = (\boldsymbol{I}+\boldsymbol{G}^2)^{-1/2}$ and $\boldsymbol{Y} = \boldsymbol{GX}$~\cite{Roh25}. For a GHZ state, we start from a star graph state and then apply $-\pi/2$ phase to the center node. 

To take into account an optical loss, we consider vacuum noise coupled by a beam splitter. Optical loss amounting to $l$ changes the covariance matrix $\boldsymbol{V}$ in Eq. (\ref{eq:pureV}) as follows:
\begin{equation}
	\boldsymbol{V}' = (1-l)\boldsymbol{V}+l \boldsymbol{I}.
\end{equation}
$\boldsymbol{V}'$ is the resulting covariance matrix, and $\boldsymbol{I}$ is the identity matrix (the vacuum covariance matrix). In Section~\ref{sec:benchmark}, we use $r = (3 \ln 10)/10$ (about 6 dB squeezing) and $l = 0.3$ for all simulations.


\begin{thebibliography}{99}
\bibitem{Larsen19}
M. V. Larsen, X. Guo, C. R. Breum, J. S. Neergaard-Nielsen, and U. L. Andersen, \textit{Deterministic generation of a two-dimensional cluster state}, \href{https://doi.org/10.1126/science.aay4354}{Science \textbf{366}, 369 (2019).}

\bibitem{Asavanant19}
W. Asavanant, Y. Shiozawa, S. Yokoyama, B. Charoen-sombutamon, H. Emura, R. N. Alexander, S. Takeda, J.-i. Yoshikawa \textit{et al.}, \textit{Generation of time-domain-multiplexed two-dimensional cluster state}, \href{https://doi.org/10.1126/science.aay2645}{Science \textbf{366}, 373 (2019).}

\bibitem{Roh25}
C. Roh, G. Gwak, Y.-D. Yoon, and Y.-S. Ra, \textit{Generation of three-dimensional entangled state}, \href{https://doi.org/10.1038/s41566-025-01631-2}{Nat. Photon. \textbf{19}, 526 (2025).}

\bibitem{Barakat2025}
I. Barakat, M. Kalash, D. Scharwald, P. Sharapova, N. Lindlein, and M. Chekhova, \textit{Simultaneous measurement of multimode squeezing through multimode phase sensitive amplification}, \href{https://doi.org/10.1364/OPTICAQ.524682}{Opt. Quantum \textbf{3}, 36 (2025).}

\bibitem{Chen2023}
W. Chen, Y. Lu, S. Zhang, K. Zhang, G. Huang, M. Qiao, X. Su, J. Zhang, J.-N. Zhang, L. Banchi \textit{et al.}, \textit{Scalable and programmable photonic network with trapped ions}, \href{https://doi.org/10.1038/s41567-023-01952-5}{Nat. Phys. \textbf{19}, 877 (2023).}

\bibitem{Li25}
Y. Li, Y. Li, X. Cheng, L. Wang, X. Zhao, W. Hou, K. Rehan, M. Zhu, L. Yan, X. Qin \textit{et al.}, \textit{Programmable multi-mode entanglement via dissipative engineering in vibrating trapped ions}, \href{https://doi.org/10.1126/sciadv.adv7838}{Sci. Adv. \textbf{11}, eadv7838 (2025).}

\bibitem{Kotler2021}
S. Kotler, G. A. Peterson, E. Shojaee, F. Lecocq, K. Cicak, A. Kwiatkowski, S. Geller, S. Glancy, E. Knill, R. W. Simmonds \textit{et al.}, \textit{Direct observation of deterministic macroscopic entanglement}, \href{https://doi.org/10.1126/science.abf2998}{Science \textbf{372}, 622 (2021).}

\bibitem{Jeong25}
M. Jeong, H. Seok, Y.-S. Ra, and J. Suh, \textit{Simultaneous generation and transfer of mechanical noise squeezing}, \href{https://doi.org/10.48550/arXiv.2509.06102}{arXiv:2509.06102.}

\bibitem{Mallet2011}
F. Mallet, M. A. Castellanos-Beltran, H. S. Ku, S. Glancy, E. Knill, K. D. Irwin, G. C. Hilton, L. R. Vale, and K. W. Lehnert, \textit{Quantum state tomography of an itinerant squeezed microwave field}, \href{https://doi.org/10.1103/PhysRevLett.106.220502}{Phys. Rev. Lett. \textbf{106}, 220502 (2011).}

\bibitem{Lingua2025}
F. Lingua, J. C. Rivera Hernández, M. Cortinovis, and D. B. Haviland, \textit{Continuous-variable square-ladder cluster states in a microwave frequency comb}, \href{https://doi.org/10.1103/PhysRevLett.134.183602}{Phys. Rev. Lett. \textbf{134}, 183602 (2025).}

\bibitem{Zhong20}
H.-S. Zhong, H. Wang, Y.-H. Deng, M.-C. Chen, L.-C. Peng, Y.-H. Luo, J. Qin, D. Wu, X. Ding, Y. Hu \textit{et al.}, \textit{Quantum computational advantage using photons}, \href{https://doi.org/10.1126/science.abe8770}{Science \textbf{370}, 1460 (2020).}

\bibitem{Madsen22}
L. S. Madsen, F. Laudenbach, M. F. Askarani, F. Rortais, T. Vincent, J. F. Bulmer, F. M. Miatto, L. Neuhaus, L. G. Helt, M. J. Collins \textit{et al.}, \textit{Quantum computational advantage with a programmable photonic processor}, \href{https://doi.org/10.1038/s41586-022-04725-x}{Nature \textbf{606}, 75 (2022).}

\bibitem{Kovalenko21}
O. Kovalenko, Y.-S. Ra, Y. Cai, V. C. Usenko, C. Fabre, N. Treps, and R. Filip, \textit{Frequency-multiplexed entanglement for continuous-variable quantum key distribution}, \href{https://doi.org/10.1364/PRJ.434979}{Photon. Res. \textbf{9}, 2351 (2021).}

\bibitem{Shi2023}
S. Shi, Y. Wang, L. Tian, W. Li, Y. Wu, Q. Wang, Y. Zheng, and K. Peng, \textit{Continuous variable quantum teleportation network}, \href{https://doi.org/10.1002/lpor.202200508}{Laser Photonics Rev. \textbf{17}, 2200508 (2023).}

\bibitem{Guo20}
X. Guo, C. R. Breum, J. Borregaard, S. Izumi, M. V. Larsen, T. Gehring, M. Christandl, J. S. Neergaard-Nielsen, and U. L. Andersen, \textit{Distributed quantum sensing in a continuous-variable entangled network}, \href{https://doi.org/10.1038/s41567-019-0743-x}{Nat. Phys. \textbf{16}, 281 (2020).}

\bibitem{Barbieri:2022hq}
M. Barbieri, \textit{Optical quantum metrology}, \href{https://doi.org/10.1103/PRXQuantum.3.010202}{PRX Quantum \textbf{3}, 010202 (2022).}

\bibitem{James01}
D. F. V. James, P. G. Kwiat, W. J. Munro, and A. G. White, \textit{Measurement of qubits}, \href{https://doi.org/10.1103/PhysRevA.64.052312}{Phys. Rev. A \textbf{64}, 052312 (2001).}

\bibitem{Lvovsky09}
A. I. Lvovsky and M. G. Raymer, \textit{Continuous-variable optical quantum-state tomography}, \href{https://doi.org/10.1103/RevModPhys.81.299}{Rev. Mod. Phys. \textbf{81}, 299 (2009).}

\bibitem{Lvovsky04}
A. I. Lvovsky, \textit{Iterative maximum-likelihood reconstruction in quantum homodyne tomography}, \href{https://doi.org/10.1088/1464-4266/6/6/014}{J. Opt. B: Quantum Semiclass. Opt. \textbf{6}, S556 (2004).}

\bibitem{Chapman22}
J. C. Chapman, J. M. Lukens, B. Qi, R. C. Pooser, and N. A. Peters, \textit{Bayesian homodyne and heterodyne tomography}, \href{https://doi.org/10.1364/OE.456597}{Opt. Express \textbf{30}, 15184 (2022).}

\bibitem{Ra:2020gg}
Y.-S. Ra, A. Dufour, M. Walschaers, C. Jacquard, T. Michel, C. Fabre, and N. Treps, \textit{Non-Gaussian quantum states of a multimode light field}, \href{https://doi.org/10.1038/s41567-019-0726-y}{Nat. Phys. \textbf{16}, 144 (2020).}

\bibitem{Gross10}
D. Gross, Y.-K. Liu, S. T. Flammia, S. Becker, and J. Eisert, \textit{Quantum state tomography via compressed sensing}, \href{https://doi.org/10.1103/PhysRevLett.105.150401}{Phys. Rev. Lett. \textbf{105}, 150401 (2010).}

\bibitem{He24}
K. He, M. Yuan, Y. Wong, S. Chakram, A. Seif, L. Jiang, and D. I. Schuster, \textit{Efficient multimode Wigner tomography}, \href{https://doi.org/10.1038/s41467-024-48573-x}{Nat. Commun. \textbf{15}, 4138 (2024).}

\bibitem{Tiunov20}
E. S. Tiunov, V. V. Tiunova, A. E. Ulanov, A. Lvovsky, and A. K. Fedorov, \textit{Experimental quantum homodyne tomography via machine learning}, \href{https://doi.org/10.1364/OPTICA.389482}{Optica \textbf{7}, 488 (2020).}

\bibitem{Hsieh:2022jc}
H.-Y. Hsieh, Y.-R. Chen, H.-C. Wu, H. L. Chen, J. Ning, Y.-C. Huang, C.-M. Wu, and R.-K. Lee, \textit{Extract the degradation information in squeezed states with machine learning}, \href{https://doi.org/10.1103/PhysRevLett.128.073604}{Phys. Rev. Lett. \textbf{128}, 073604 (2022).}

\bibitem{Weedbrook12}
C. Weedbrook, S. Pirandola, R. Garc\'{\i}a-Patr\'on, N. J. Cerf, T. C. Ralph, J. H. Shapiro,  and S. Lloyd, \textit{Gaussian quantum information}, \href{https://doi.org/10.1103/RevModPhys.84.621}{Rev. Mod. Phys. \textbf{84}, 621 (2012).}

\bibitem{Serafini2017}
A. Serafini, \textit{Quantum Continuous Variables: A Primer of Theoretical Methods} (CRC Press, Boca Raton, 2017).

\bibitem{Mele25}
F. A. Mele, A. A. Mele, L. Bittel, J. Eisert, V. Giovannetti, L. Lami, L. Leone, and S. F. Oliviero, \textit{Learning quantum states of continuous-variable systems}, \href{https://doi.org/10.1038/s41567-025-03086-2}{Nat. Phys. \textbf{21}, 2002 (2025).}

\bibitem{D'Auria09}
V. D'Auria, S. Fornaro, A. Porzio, S. Solimeno, S. Olivares, and M. G. A. Paris, \textit{Full characterization of Gaussian bipartite entangled states by a single homodyne detector}, \href{https://doi.org/10.1103/PhysRevLett.102.020502}{Phys. Rev. Lett. \textbf{102}, 020502 (2009).}

\bibitem{Roslund14}
J. Roslund, R. M. de Araújo, S. Jiang, C. Fabre, and N. Treps, \textit{Wavelength-multiplexed quantum networks with ultrafast frequency combs}, \href{https://doi.org/10.1038/nphoton.2013.340}{Nat. Photon. \textbf{8}, 109 (2014).}

\bibitem{Cai17}
Y. Cai, J. Roslund, G. Ferrini, F. Arzani, X. Xu, C. Fabre, and N. Treps, \textit{Multimode entanglement in reconfigurable graph states using optical frequency combs}, \href{https://doi.org/10.1038/ncomms15645}{Nat. Commun. \textbf{8}, 15645 (2017).}

\bibitem{Banchi15}
L. Banchi, S. L. Braunstein, and S. Pirandola, \textit{Quantum fidelity for arbitrary Gaussian states}, \href{https://doi.org/10.1103/PhysRevLett.115.260501}{Phys. Rev. Lett. \textbf{115}, 260501 (2015).}

\bibitem{Simon00}
R. Simon, \textit{Peres-Horodecki separability criterion for continuous variable systems}, \href{https://doi.org/10.1103/PhysRevLett.84.2726}{Phys. Rev. Lett. \textbf{84}, 2726 (2000).}

\bibitem{Kogias15}
I. Kogias, A. R. Lee, S. Ragy, and G. Adesso, \textit{Quantification of Gaussian quantum steering}, \href{https://doi.org/10.1103/PhysRevLett.114.060403}{Phys. Rev. Lett. \textbf{114}, 060403 (2015).}

\bibitem{Gerke15}
S. Gerke, J. Sperling, W. Vogel, Y. Cai, J. Roslund, N. Treps, and C. Fabre,\textit{Full multipartite entanglement of frequency-comb Gaussian states}, \href{https://doi.org/10.1103/PhysRevLett.114.050501}{Phys. Rev. Lett. \textbf{114}, 050501 (2015).}

\bibitem{Shchukin24}
E. Shchukin and P. van Loock, \textit{Revisiting Gaussian genuine entanglement witnesses with modern software}, \href{https://doi.org/10.48550/arXiv.2412.09757}{arXiv:2412.09757 (2024).}

\bibitem{Gwak25}
G. Gwak, C. Roh, Y.-D. Yoon, M. S. Kim, and Y.-S. Ra, \textit{Completely characterizing multimode second-order nonlinear optical quantum processes}, \href{https://doi.org/10.1038/s41566-025-01787-x}{Nat. Photon. \textbf{20}, 156 (2026).}

\bibitem{Williamson36}
J. Williamson, \textit{On the algebraic problem concerning the normal forms of linear dynamical systems}, \href{https://doi.org/10.2307/2371062}{Am. J. Math. \textbf{58}, 141 (1936).}

\bibitem{Bloch62}
C. Bloch and A. Messiah, \textit{The canonical form of an antisymmetric tensor and its application to the theory of superconductivity}, \href{https://doi.org/10.1016/0029-5582(62)90377-2}{Nucl. Phys. \textbf{39}, 95 (1962).}

\bibitem{Raymer99}
M. G. Raymer and A. C. Funk, \textit{Quantum-state tomography of two-mode light using generalized rotations in phase space}, \href{https://doi.org/10.1103/PhysRevA.61.015801}{Phys. Rev. A \textbf{61}, 015801 (1999).}

\bibitem{Schwemmer2015}
C. Schwemmer, L. Knips, D. Richart, H. Weinfurter, T. Moroder, M. Kleinmann, and O. Gühne, \textit{Systematic errors in current quantum state tomography tools}, \href{https://doi.org/10.1103/PhysRevLett.114.080403}{Phys. Rev. Lett. \textbf{114}, 080403 (2015).}

\bibitem{Aditi2025}
K. Aditi and S. Becker, \textit{Rigorous maximum-likelihood estimation for quantum states}, \href{https://doi.org/10.1103/j5gh-hmtw}{Phys. Rev. A \textbf{112}, 052436 (2025).}

\bibitem{Hradil1997}
Z. Hradil, \textit{Quantum-state estimation}, \href{https://doi.org/10.1103/PhysRevA.55.R1561}{Phys. Rev. A \textbf{55}, R1561 (1997).}

\bibitem{Petrov20}
A. G. Petrov, \textit{On parametric representations of orthogonal and symplectic matrices}, \href{https://doi.org/10.3103/S1066369X20060122}{Russ. Math. \textbf{64}, 80 (2020).}

\bibitem{Kingma14}
D. P. Kingma and J. Ba, \textit{Adam: A method for stochastic optimization}, \href{https://doi.org/10.48550/arXiv.1412.6980}{arXiv:1412.6980 (2014).}

\bibitem{vanLoock2007}
P. van Loock, C. Weedbrook, and M. Gu, \textit{Building Gaussian cluster states by linear optics}, \href{https://doi.org/10.1103/PhysRevA.76.032321}{Phys. Rev. A. \textbf{76}, 032321 (2007).}

\bibitem{Loock00}
P. van Loock and S. L. Braunstein, \textit{Multipartite entanglement for continuous variables: A quantum teleportation network}, \href{https://doi.org/10.1103/PhysRevLett.84.3482}{Phys. Rev. Lett. \textbf{84}, 3482 (2000).}

\bibitem{Loock15}
E. Shchukin and P. van Loock, \textit{Generalized conditions for genuine multipartite continuous-variable entanglement}, \href{https://doi.org/10.1103/PhysRevA.92.042328}{Phys. Rev. A \textbf{92}, 042328 (2015).}

\bibitem{Jiang14}
Z. Jiang, \textit{Quantum Fisher information for states in exponential form}, \href{https://doi.org/10.1103/PhysRevA.89.032128}{Phys. Rev. A. \textbf{89}, 032128 (2014).}

\bibitem{Gebhart23}
V. Gebhart, R. Santagati, A. A. Gentile, E. M. Gauger, D. Craig, N. Ares, L. Banchi, F. Marquardt, L. Pezze, and C. Bonato, \textit{Learning quantum system}, \href{https://doi.org/10.1038/s42254-022-00552-1}{Nat. Rev. Phys. \textbf{5}, 141 (2023).}

\bibitem{Ra:2017ia}
Y.-S. Ra, C. Jacquard, A. Dufour, C. Fabre, and N. Treps, \textit{Tomography of a mode-tunable coherent single photon subtractor}, \href{https://doi.org/10.1103/PhysRevX.7.031012}{Phys. Rev. X \textbf{7}, 031012 (2017).}

\bibitem{Anshu2023}
A. Anshu and S. Arunachalam, \textit{A survey on the complexity of learning quantum states}, \href{https://doi.org/10.1038/s42254-023-00662-4}{Nat. Rev. Phys. \textbf{6}, 59 (2024).}

\end{thebibliography}


\end{document}